\documentclass[english,aps,twocolumn]{revtex4-1}
\usepackage[T1]{fontenc}
\usepackage[latin9]{inputenc}
\usepackage{babel}
\usepackage{amsmath}
\usepackage{amssymb}
\usepackage{graphicx}
\usepackage[unicode=true,pdfusetitle,breaklinks=false,pdfborder={0 0 1},backref=false,colorlinks=false]{hyperref}

\newcommand{\avgR}{\left\langle \tilde{r}\right\rangle}

\newcommand{\TwoGOE}{{$2\times$GOE}}
\newcommand{\TwoGUE}{{$2\times$GUE}}
\newcommand{\TwoGSE}{{$2\times$GSE}}
\newcommand{\TwoGxE}{{$2\times$G$x$E}}

\makeatletter

\providecommand{\tabularnewline}{\\}

\makeatother

\begin{document}
\title{Exact gap-ratio results for mixed Wigner surmises of up to 4 eigenvalues}
\author{Mikael Fremling}
\affiliation{Institute for Theoretical Physics and Center for Extreme Matter and
Emergent Phenomena, Utrecht University, Princetonplein 5, 3584 CC
Utrecht, The Netherlands }
\begin{abstract}
  We compute some exact results for the gap-ratio of mixed Wigner surmises for up to four eigenvalues and $0\leq\beta\leq4$.
  The main results concern equal mixtures of the GOE, GUE, and GSE random matrix classes. These give rise to \TwoGOE, \TwoGUE, and \TwoGSE~distributions.
  We find that \TwoGOE, \TwoGUE~are well approximated by the surmises of only 2+2 eigenvalues that are GOE and GUE distributed, respectively.
  The same is not valid for \TwoGSE, which is well estimated, by coincidence,
  by 2+2 eigenvalues of statistics intermediate between GUE and GSE.
\end{abstract}
\maketitle

\section{Introduction}

Random matrix theory (RMT) can be used to analyze the spectral statistics of quantum mechanical Hamiltonians.
If a Hamiltonian is non-integrable or has a chaotic semiclassical limit,
then its distribution of energy levels will generically follow one of the three random matrix ensembles Gaussian-Orthogonal-Ensemble (GOE),
Gaussian-Unitary-Ensemble (GUE), or Gaussian-Symplectic-Ensemble (GSE).
If the Hamiltonian is integrable, the energy levels appear uncorrelated,
resulting in Poisson statistics\cite{Berry1977, Bohigas1984, Casati1980, Berry1985}.

\begin{figure}
\begin{centering}
  \includegraphics[width=1.0\linewidth]{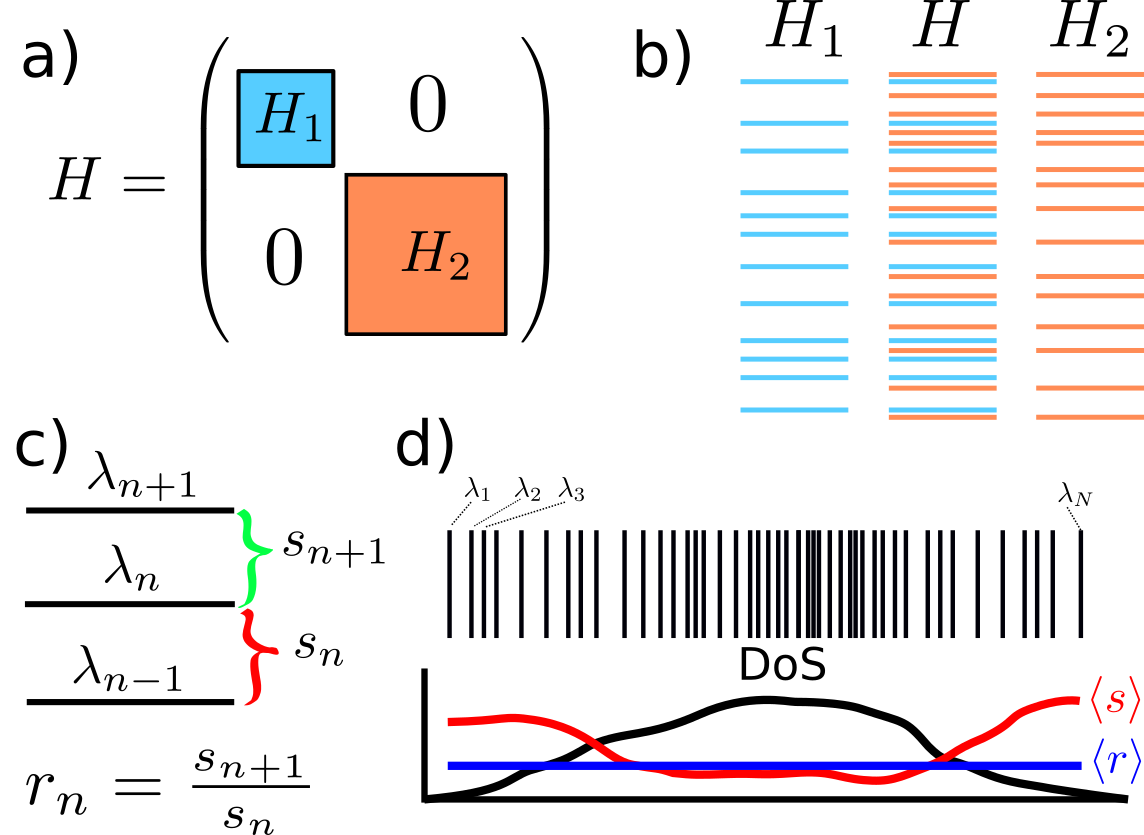}
  \par\end{centering}
  \caption{a) If a Hamiltonian, $H$, contains a symmetry, it can be used to split the Hamiltonian into two (or more) blocks $H_1$, $H_2$, ... whose spectra are independent. b) The spectra of $H_1$ and $H_2$ will show level-repulsion, which is masked in $H$ where the spectra are superimposed.
    c) Pictorial representation of the level spacing, $s$, and the gap-ratio, $r$.
    d) In a typical Hamiltonian (random matrix), the density of states (DoS) is not constant at all energies.
    As a consequence, the ``local'' gap spacing, $\left<s\right>$, is not constant, which can be remedied by unfolding.
    On the other hand, the ``local'' gap-ratio, $\left<r\right>$, is constant, making it easier to work with it.
\label{fig_intro}}
\end{figure}

We can refine the quite generic picture described above by considering Hamiltonians with a single, or a few, unresolved symmetries.
An unresolved symmetry is a quantum number that block-diagonalizes the Hamiltonian, but which has not been utilized.
As a result, the spectral statistics will break into two (or more) independent ensembles.
If the spectrum breaks into two independent GOE, GUE, or GSE ensembles,
we call them \TwoGOE, \TwoGUE, and \TwoGSE, respectively.
The \TwoGxE~($x=$O,U,S) ensembles are not the same as the Poissonian of independent variables, as not all eigenvalues are uncorrelated.
For an illustration of this process, see Fig.~\ref{fig_intro}a-b).

Many situations could lead to the \TwoGxE~ensembles listed above, for instance,
when there is an unresolved $\mathbb{Z}_{2}$ symmetry. In the recent literature,
one can find examples in Floquet systems \cite{Gazit2018, Chan2018, Khemani2016}, the Fractional Quantum Hall Effect \cite{Fremling2018},
restricted SYK models \cite{Milekhin2021,Fremling2021a,Fremling2021b},
and fracton models \cite{Pretko2020, Sala2020, Khemani2020, Moudgalya2019}.

To study the spectral statistics of a Hamiltonian with (ordered) eigenvalues $\lambda_n$,
an early method was that of the nearest-neighbor spacing distribution for the variable $s_n=\lambda_{n}-\lambda_{n-1}$.
However, the distance energy spacing depends on the overall energy scale which needs to be divided away.
Further, the local density of states could depend on energy, which would be accounted for with a process called unfolding, see Fig.~\ref{fig_intro}c-d)

To circumvent the somewhat arbitrary unfolding procedure, Oganesyan and Huse \cite{Oganesyan2007} suggested using the gap-ratio distribution
\[r_n=\frac{s_{n+1}}{s_n}=\frac{\lambda_{n+1}-\lambda_{n}}{\lambda_{n}-\lambda_{n-1}},\]
since it is invariant under smooth scale changes.
In two papers, Atas \emph{et al.} \cite{Atas2013,Atas2013b} derived a set of analytic results (see Eqn.~\eqref{eq_Attas_result}) regarding the distribution of distribution of $r$,
based on the Wigner surmise for a small number of eigenvalues.

\begin{table*}[t]
\begin{tabular}{|ccc|ccc|ccc|}
\hline 
\multicolumn{1}{|c|}{Distributions} & \multicolumn{1}{c|}{Eqn.~no.} & $\left\langle r\right\rangle $ & \multicolumn{1}{c|}{Distribution} & \multicolumn{1}{c|}{Eqn.~no.} & $\left\langle r\right\rangle $ & \multicolumn{1}{c|}{Distribution} & \multicolumn{1}{c|}{Eqn.~no.} & $\left\langle r\right\rangle $\tabularnewline
\hline 
$P_{1,1}^{2,2}(r)$ & (\ref{eq_P^(2,2)_(1,1)}) & 0.422798 & $P_{1,0}^{2,1}(r)$ & (\ref{eq_P^(2,1)_(1,0)}) & 0.419427 & $P_{0,1}^{2,2}(r)$ & \ref{eq_P^(2,2)_(0,1)} & 0.403566\tabularnewline
$P_{2,2}^{2,2}(r)$ & (\ref{eq_P^(2,2)_(2,2)}) & 0.420518 & $P_{2,0}^{2,1}(r)$ & (\ref{eq_P^(2,1)_(2,0)}) & 0.408545 & $P_{0,2}^{2,2}(r)$ & \ref{eq_P^(2,2)_(0,2)} & 0.398237\tabularnewline
$P_{3,3}^{2,2}(r)$ & (\ref{eq_P^(2,2)_(3,3)}) & 0.409623 & $P_{3,0}^{2,1}(r)$ & (\ref{eq_P^(2,1)_(3,0)}) & 0.404868 & $P_{0,3}^{2,2}(r)$ & \ref{eq_P^(2,2)_(0,3)} & 0.395907\tabularnewline
$P_{4,4}^{2,2}(r)$ & (\ref{eq_P^(2,2)_(4,4)}) & 0.39371 & $P_{4,0}^{2,1}(r)$ & (\ref{eq_P^(2,1)_(4,0)}) & 0.408545 & $P_{0,4}^{2,2}(r)$ & \ref{eq_P^(2,2)_(0,4)} & 0.395762\tabularnewline
\hline 
\multicolumn{1}{|c|}{Distribution} & \multicolumn{1}{c|}{Eqn.~no.} & \multicolumn{1}{c||}{$\left\langle r\right\rangle $} & $P_{0}^{4}(r)$ & (\ref{eq_P^4_0}) & \multicolumn{1}{c}{0.398237} & $P_{1}^{4}(r)$ & (\ref{eq_P^4_1}) & 0.531785\tabularnewline
\hline 
$P_{1,2}^{2,2}(r)$ & (\ref{eq_P^(2,2)_(1,2)}) & 0.423367 & $P_{2,3}^{2,2}(r)$ & \ref{eq_P^(2,2)_(2,3)} & 0.418542 & $P_{1,0}^{3,1}(r)$ & (\ref{eq_P^(3,1)_(1,0)}) & 0.429718\tabularnewline
$P_{1,3}^{2,2}(r)$ & \ref{eq_P^(2,2)_(1,3)} & 0.426026 & $P_{2,4}^{2,2}(r)$ & \ref{eq_P^(2,2)_(2,4)} & 0.420494 & $P_{2,0}^{3,1}(r)$ & (\ref{eq_P^(3,1)_(2,0)}) & 0.420664\tabularnewline
$P_{1,4}^{2,2}(r)$ & \ref{eq_P^(2,2)_(1,4)} & 0.431454 & $P_{3,4}^{2,2}(r)$ & \ref{eq_P^(2,2)_(3,4)} & 0.405069 & $P_{3,0}^{3,1}(r)$ & (\ref{eq_P^(3,1)_(3,0)}) & 0.40585\tabularnewline
\hline 
\end{tabular}

\caption{Table of the various mixed Wigner surmises considered in this work,
their equation number and expectation value for $\tilde{r}_{n}$ in
Eqn.~(\ref{eq_r_tilde}).\label{tab:Results}}
\end{table*}

In recent years, a number of generalizations have been added to the literature on higher-order level spacings
\cite{Chavda2013, Bhosale2018, Tekur2018, Tekur2018b, Tekur2020, Giraud2022, Corps2020, Corps2021}
and non-hermitian Hamiltonians \cite{Sa2020}.
This short paper adds to the current literature by computing exact $r$-distribution functions for mixed Wigner surmises, such as two superimposed GUE spectra. In passing,
we will also collect some lesser-known results from the literature.
The work in Ref.~\onlinecite{Giraud2022} also considers level statistics of mixed ensembles,
but from a slightly different perspective than the present work.

Computer algebra has been used extensively to minimize clerical errors when deriving the results of this paper.
We use custom-built Julia code that communicates with Mathematica by using the MathLink and MathLinkExtras packages,
and the implementation is outlined in Appendix \ref{app_Recursive-integration}.

This paper is organized as follows:
In Section \ref{sec: The-Wigner-Surmise} we introduce some notation,
and recapitulate the result of Atas \emph{et al.} \cite{Atas2013}.
We also discuss the added complications independent (mixed) Wigner surmises.
In section \ref{sec:Numerics} we perform numerical tests,
while the functional form of our analytical results is presented in section \ref{sec:Analytical-Results}.
As a reference, the main results, and their equation numbers, are summarized in Table \ref{tab:Results}.

\begin{figure*}
\begin{centering}
\includegraphics[width=1.0\linewidth]{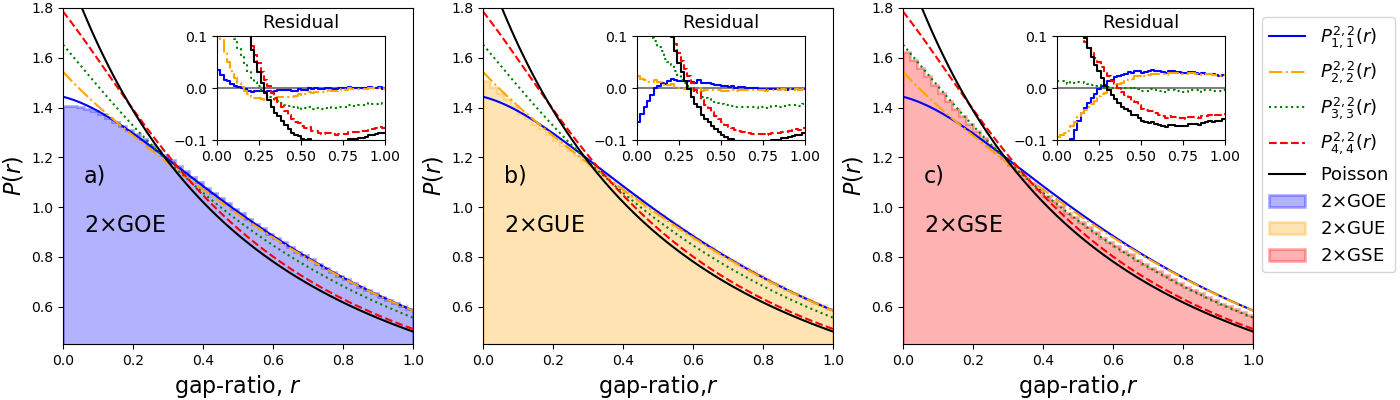}
\par\end{centering}
\caption{Comparison of the asymptotic \TwoGxE distributions to the analytic $2+2$ surmises $P_{\beta,\beta}^{2,2}(r)$, for $\beta=1,2,3,4$.
The inset shows the residual $P_{\beta,\beta}^{2,2} -P_{2\times\mathrm{G}x\mathrm{E}}$.
For a) \TwoGOE~and b) \TwoGUE, the surmises $P_{1,1}^{2,2}$ and $P_{2,2}^{2,2}$ give a good approximation,
with some deviation at small $r$.
On the other hand,  \TwoGSE~is not approximated well with the surmise $P_{4,4}^{2,2}$.
However, by coincidence, $P_{3,3}^{2,2}$ gives an excellent surmise for \TwoGSE.\label{fig_Comparison}}
\end{figure*}

\section{The Wigner Surmise \label{sec: The-Wigner-Surmise}}

In this section we begin by reviewing the approach taken by Ref.~\onlinecite{Atas2013} to compute the gap-ratio distribution $P_{\beta}^{3}(r)$.
As starting point we will use the Wigner surmises for the Random Matrix ensembles G$x$E with $x=\text{O},\text{U},\text{S}$,
which is known to approximate well the statistics for large random matrices \cite{Dietz1990}.
For $N$ eigenvalues the surmise takes the form
\begin{equation}
P_\beta^N\left(\lambda\right)\propto e^{-\sum_{j=1}^N\lambda_j^2}\prod_{i<j=1}^{N}\left|\lambda_i-\lambda_j\right|^\beta\label{eq_WignerSurmise}
\end{equation}
 where $\beta=1,2,4$ corresponding to $x=\text{O},\text{U},\text{S}$.
This form is exact for $2\times2$ matrices and also agrees with the asymptotic distribution \cite{Dietz1990}.
From $P_{\beta}^{N}\left(\lambda\right)$ one computes the $r$-statistics by sorting the eigenvalues $\lambda$ and computing the distribution of
\[r_{n}=\frac{\lambda_{n+1}-\lambda_{n}}{\lambda_{n}-\lambda_{n-1}}.\]
In this work, just like in Ref.~\onlinecite{Atas2013}, we define $r$ on the semi-open interval $r\in\left[0,\infty\right)$ rather than the closed interval $\tilde{r}\in\left[0,1\right]$ corresponding to
  \begin{equation}
    \tilde{r}_n=\text{min}\left(r_n,1/r_n\right)\label{eq_r_tilde}
    \end{equation}
In principle $P(r)$ and $P(\tilde r)$ could be different, however, if $P(r)$ has the property $P\left(\frac{1}{r}\right)=P(r)r^{2}$ then $P(\tilde r)=2P(r)$.
This is the case for all distributions considered here.

For $N=3$ one may, without loss of generality, assume that $-\infty<\lambda_{1}<\lambda_{2}<\lambda_{3}<\infty$ and (\ref{eq_WignerSurmise}) takes the form of

\begin{align}
 & P_{\beta}^{3}\left(\lambda_{1},\lambda_{2},\lambda_{3}\right)\propto e^{-\sum_{n=1}^{3}\lambda_{n}^{2}}\label{eq_P_3}\\
 & \phantom{aaaaaaa}\times\left(\lambda_{2}-\lambda_{1}\right)^{\beta}\left(\lambda_{3}-\lambda_{1}\right)^{\beta}\left(\lambda_{3}-\lambda_{2}\right)^{\beta}.\nonumber 
\end{align}
 The distribution for $r$ is computed by multiplying with $\delta\left(r-\frac{\lambda_{3}-\lambda_{2}}{\lambda_{2}-\lambda_{1}}\right)$ and performing the nested integrals over $\lambda_{1},\lambda_{2},\lambda_{3}$ as 
\begin{align}
 & P_{\beta}^{3}(r)\propto\int_{-\infty}^{\infty}d\lambda_{1}\,\int_{\lambda_{1}}^{\infty}d\lambda_{2}\,\int_{\lambda_{2}}^{\infty}d\lambda_{3}\label{eq_P(r)}\\
 & \phantom{aaaaaaa}\times P_{\beta}^{3}\left(\lambda_{1},\lambda_{2},\lambda_{3}\right)\,\delta\left(r-\frac{\lambda_{3}-\lambda_{2}}{\lambda_{2}-\lambda_{1}}\right)\nonumber 
\end{align}
where the delta function $\delta\left(r-\frac{\lambda_{3}-\lambda_{2}}{\lambda_{2}-\lambda_{1}}\right)$ is inserted to ensure that $\int_{0}^{\infty}dr\,P_{\beta}^{3}(r)=1$.
The first integral can easily be evaluated to give 
\begin{align*}
P_{\beta}^{3}(r) & \propto\left(r^{2}+r\right)^{\beta}\int_{-\infty}^{\infty}d\lambda_{1}\,\int_{\lambda_{1}}^{\infty}d\lambda_{2}\,\\
 & \phantom{aaaa}\times e^{-\lambda_{1}^{2}-\lambda_{2}^{2}-\left(\lambda_{2}+r\left(\lambda_{2}-\lambda_{1}\right)\right)^{2}}\left(\lambda_{2}-\lambda_{1}\right)^{3\beta+1}.
\end{align*}

Further by rewriting $\lambda_{2}=\delta+\lambda_{1}$ the integral over $\lambda_{1}$ can immediately performed to give 
\begin{align*}
P_{\beta}^{3}(r) & \propto\left(r^{2}+r\right)^{\beta}\,\int_{0}^{\infty}d\delta\,e^{-\frac{2}{3}\delta^{2}\left(r^{2}+r+1\right)}\delta^{3\beta+1}.
\end{align*}
 The final Gaussian integral is readily evaluated to yield the result
\begin{equation}
P_{\beta}^{3}(r)=\frac{1}{Z_{\beta}}\frac{\left(r+r^{2}\right)^{\beta}}{\left(1+r+r^{2}\right)^{1+\beta\frac{3}{2}}},\label{eq_Attas_result}
\end{equation}
 with $Z_{\beta}$ being the normalization constant. 

\subsection{Mixed Surmises}

In this work we will focus on products of Wigner surmises that allow for uncorrelated eigenvalues. Equation (\ref{eq_WignerSurmise}) is then generalized as 
\begin{equation}
P_{\vec{\beta}}^{\vec{N}}\left(\lambda\right)\propto\prod_{\alpha}P_{\beta_{\alpha}}^{N_{\alpha}}\left(\lambda\in\Lambda_{\alpha}\right),\label{eq_WignerSurmise-Mix}
\end{equation}
where $\Lambda_{\alpha}$ is the set of $N_{\alpha}$ eigenvalues that follow the Wigner surmise with coefficient $\beta_{\alpha}$.
Distributions of this form are found in the spectrum when there are unresolved symmetries in the Hamiltonian.

Compared with Eqn.~(\ref{eq_WignerSurmise}),
there are also two extra levels of complication when computing $P_{\vec{\beta}}^{\vec{N}}(r)$.
The first is that one may not assume that $\lambda_{i}>\lambda_{j}$ if $i$ and $j$ come from different sets of eigenvalues.
Rather one needs to separately treat all permutations of $N=\sum_{\alpha}N_{\alpha}$ elements divided into groups $\Lambda_{\alpha}$ of sizes $N_{\alpha}$.
The multi-nomial coefficient
\[ M=\frac{N!}{\prod_{\alpha}N_{\alpha}!},\]
then gives the number $M$ of such combinations.
Each such permutation comes with a different ordering of the $\lambda$s, and they need all be taken into account.
Secondly, for a given permutation there are $N-2$ different gap ratios that contribute to the final distribution $P_{\vec{\beta}}^{\vec{N}}(r)$.

Thus, for a given ordering $\sigma$ of the eigenvalues $\lambda$, we may define a map $g_{j}=\lambda_{\sigma\left(j\right)}$ to a new set of variables $g$.
This set will have the property that $g_{i}>g_{j}$ if (and only if) $i>j$.
The generalization of Eq.~\eqref{eq_P(r)} for $M$ permutations of $N$ eigenvalues then reads
\[P_{\vec{\beta}}^{\vec{N}}(r)\propto\sum_{\sigma=1}^{M}\sum_{k=1}^{N-2}P_{k,\sigma}(r).\]
Here, $P_{k,\sigma}(r)$ is the probability distribution for the $k$'th gap ratio of the permutation $\sigma$,
and $k=1,\ldots,N-2$ enumerates the different gap rations averaged over.
It takes the form
\begin{align}
P_{k,\sigma}(r) & =\int_{-\infty}^{\infty}dg_{1}\,\int_{g_{1}}^{\infty}dg_{2}\,\ldots\int_{g_{N-1}}^{\infty}dg_{N}\nonumber \\
 & \phantom{aaaa}\times\,P_{\sigma}\left(g\right)\,\delta\left(r-\frac{g_{k+2}-g_{k+1}}{g_{k+1}-g_{k}}\right),\label{eq_integral_term}
\end{align}
 where $P_{\sigma}\left(g\right)=P_{\vec{\beta}}^{\vec{N}}\left(\lambda_{\sigma\left(j\right)}\right)$ is the (unnormalized) conditional probability distribution for the permutation $\sigma$.
Needless to say, the scope for errors in algebraic bookkeeping is paramount.

In the present work, we solve both of these combinatorial headaches by automating the integration procedure and using computer-generated Mathematica and Julia code.
For a more detailed discussion of the setup, we refer to Appendix \ref{app_Recursive-integration},
and in the following section, we merely present the results.

\section{Comparison to numerics\label{sec:Numerics}}

To gauge the usefulness of these new distributions, we focus on the $P_{\beta,\beta}^{2,2}(r)$ distributions and compare with numerical results.
To obtain the asymptotic \TwoGxE~distribution, we constrict two independent Hermitian matrices with random entries and diagonalize them to get their real spectrum.
For $\beta=1$ ($\beta=2$), the matrices are $N\times N$ with random normal distributed real (complex) numbers.
For the GSE case, we use a $2N\times2N$ dimensional matrix with entries $q=\sum_{j=0}^{3}a_{j}k_{j}$ where the $a_{j}$ are normal-distributed.
The quaternions $k_{j}$ are represented by the $2\times2$ matrices $k_{0}=1_{2\times2}$, $k_{j}=\imath\sigma_{j}$, where $\sigma_{j}$ are the Pauli matrices.
The GSE-spectrum then has an exact two-fold degeneracy that we remove before computing the $r$-statistics. 

In Figure \ref{fig_Comparison} we first compare $P_{\beta,\beta}^{2,2}(r)$ for $\beta=1,2,3,4$ to the asymptotic distributions obtained numerically.
In the figure $P_{1,1}^{2,2}(r)$ and $P_{2,2}^{2,2}(r)$ give a reasonable surmise for \TwoGOE~and \TwoGUE, respectively.
On the other hand $P_{4,4}^{2,2}(r)$ does not give a good approximation for \TwoGSE.
However, it seems that $P_{3,3}^{2,2}(r)$ does give a good surmise for \TwoGSE.
We will see below in Fig.~\ref{fig_Scaling_Surmise}b that this is a coincidence and that the asymptotic limit of $P_{3,3}^{N,N}(r)$ gives the wrong surmise. 

In Figure \ref{fig_Scaling_Surmise}b, the coincidental nature of the good fit for \TwoGSE~using $P_{3,3}^{2,2}(r)$ is made more explicit.
In the figure, we show $\avgR$ for $P_{\beta,\beta}^{N,N}$ and compare with \TwoGxE~(dash-dotted lines).
We use standard Metropolis-Hasting Monte-Carlo sampling to generate the $P_{\beta,\beta}^{N,N}$ distribution and from there compute $\avgR$. As a guide to the eye, the analytical results of $P_{\beta,\beta}^{2,2}$ are displayed in dashed lines.
We see in the figure that the average of $P_{3,3}^{2,2}$ ($\avgR=0.4096$,
green) by accident comes close to the asymptotic 2$\times$GSE result ($\avgR=0.4114$). However, as $N$ is increased, $P_{3,3}^{N,N}$ overshoots and approaches the much higher average $\avgR=0.4166$.

\begin{figure}
\begin{centering}
\includegraphics[width=1.0\linewidth]{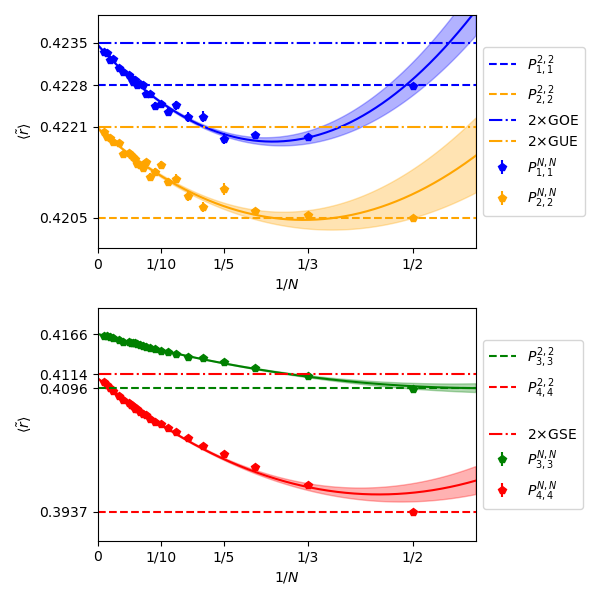}
\par\end{centering}
\caption{Comparison of $\avgR$ for $P_{\beta,\beta}^{N,N}$
  and \TwoGxE~(dash-dotted lines) for a) $\beta=1,2$ and b) $\beta=3,4$.
  As a guide to the eye the $P_{\beta,\beta}^{2,2}$ results are shown in dashed lines.
In panel a) we find, for $\beta=1$, that the surmise of $P_{1,1}^{2,2}$ is initially the best approximation of \TwoGOE~and is only improved upon at around $N=15$.
For $\beta=2$ and $\beta=3$, the improvement seems to be monotonic.
It is worth noting that the fit for \TwoGSE~by $P_{3,3}^{2,2}$ is only surpassed by $P_{4,4}^{N,N}$ at around $N=100$.
Each data point is computed from $10^{6}$ MC samples, except for $N=2,3,4$ where $10^{7}$ samples where used, to improve accuracy.
\label{fig_Scaling_Surmise}}
\end{figure}

We expect that series of surmises given by $P_{4,4}^{N,N}$ will become exact for large $N$.
Indeed, we see that $\avgR$ start too low ($\avgR=0.3937$) but then grows monotonously.
However t is only at around $N\approx100$ that $P_{4,4}^{N,N}$ comes closer than $P_{3,3}^{2,2}$.
Similar monotonic improvement with $N$ is also found for $\beta=2$.

\begin{figure}
\begin{centering}
\includegraphics[width=1.0\linewidth]{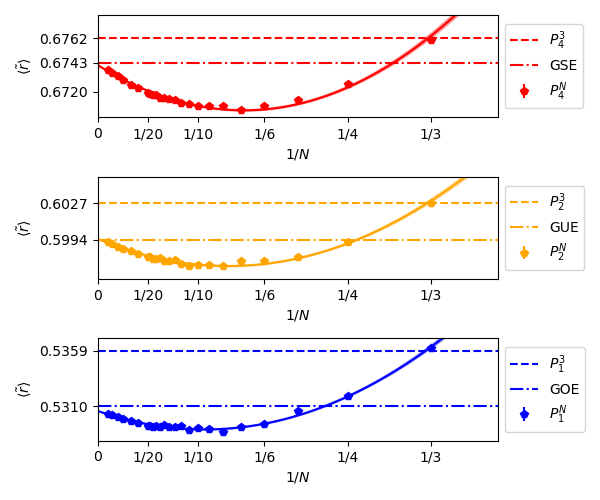}
\par\end{centering}
\caption{Comparison of $\left\langle \tilde{r}\right\rangle $ for $P_{\beta}^{N}$ and G$x$E (dash-dotted lines) for $\beta=1,2,3$.
As a guide to the eye, the $P_{\beta}^{3}$ results are shown in dashed lines.
In all three panels we find that $P_{\beta}^{3}$ initially overestimates $\left\langle \tilde{r}\right\rangle $,
and that larger $N$ brings the estimate down to then underestimate
$\left\langle \tilde{r}\right\rangle $. Finally for $N\gtrsim100$,
$P_{\beta}^{N}$ is a gain a better estimate than $P_{\beta}^{4}$.
Each data point is computed from $10^{6}$ MC samples. \label{fig_Comparison-GxE-single}}
\end{figure}

However, for $\beta=1$, the surmise $P_{1,1}^{2,2}$ remains the best until $N\approx15$.
This situation is quite similar to that of the one-component surmises $P_{\beta}^{N}$.
In that situation, there is an initial improvement for small $N$,
which overshoots and finally improves again at $N\ge100$\cite{Atas2013b}.
For clarity, this behavior is reproduced in Figure \ref{fig_Comparison-GxE-single}.

For the other surmises of the form $P_{\beta,\beta^{\prime}}^{4-N,N}$ no numerical comparisons are done,
as it is unclear that thermodynamic distributions they would approximate. 

\section{Analytical Results}\label{sec:Analytical-Results}

In this section we summarize our analytical results.
The main results if that of $P_{\beta,\beta}^{2,2}(r)$ but for completeness we also list more distributions.
As the formulas are computer generated automatically, step by step derivations will not be shown,
but intermediates results have been set aside by the algorithms,

At the moment memory restrains in prevent us from also computing $P_{\beta}^{4}(r)$ for higer values of $\beta$, but there
is no conceptual problem in going further.

To keep this section more compact we use the following abbreviations

\begin{align}
\delta & =r+1\nonumber \\
\gamma & =r^{2}+r+1\nonumber \\
d_{443} & =4r^{2}+4r+3\label{eq_abrevs}\\
d_{344} & =3r^{2}+4r+4\nonumber \\
T(r) & =\tan^{-1}\left(\frac{r+2}{2\sqrt{2\gamma}}\right)\\
T\left(\frac{1}{r}\right) & =\tan^{-1}\left(\frac{2r+1}{2\sqrt{2\gamma}}\right)
\end{align}
We also make use of the polynomials $F_{\vec{\beta}}^{\vec{N}}(r),$$G_{\vec{\beta}}^{\vec{N}}(r)$
and $E_{\vec{\beta}}^{\vec{N}}(r)$. The first two polynomials
are dual, such that $F_{\vec{\beta}}^{\vec{N}}\left(\frac{1}{r}\right)r^{\deg\left(F_{\vec{\beta}}^{\vec{N}}\right)}=G_{\vec{\beta}}^{\vec{N}}(r)$
whereas $E_{\vec{\beta}}^{\vec{N}}\left(\frac{1}{r}\right)r^{\deg\left(E_{\vec{\beta}}^{\vec{N}}\right)}=E_{\vec{\beta}}^{\vec{N}}(r)$
is self-dual.

\subsection{One component mixtures}
The one-component mixtures $P_\beta^N$ for $N=3$ where solkved for abitary integer $\beta$ in Ref.~\onlinecite{Atas2013}.
In a subsequent work\cite{Atas2013b}  they also computed the next order for $\beta=1$.
Using our method we can easilly reproduce this result
\begin{equation}
P_{1}^{4}(r)=\frac{r\delta}{4\gamma^{4}}\left(\frac{F_{1}^{4}(r)}{d_{443}^{5/2}}+\frac{G_{1}^{4}(r)}{d_{344}^{5/2}}\right)\label{eq_P^4_1}
\end{equation}
 with the auxiliary polynomials
\begin{align*}
F_{1}^{4}(r) & =512r^{8}+2048r^{7}+3768r^{6}+4136r^{5}\\
 & +2696r^{4}+888r^{3}-49r^{2}-145r-30\\
G_{1}^{4}(r) & =-30r^{8}-145r^{7}-49r^{6}+888r^{5}+2696r^{4}\\
 & +4136r^{3}+3768r^{2}+2048r+512.
\end{align*}
Ideally we would like to also target larger values of both $N$ and $\beta$ but at present this requires improved algorithms.

\subsection{Mixtures with independent level ($\beta=0$)}
We begin with listing mixtures containing uncorrelated eigenvalues ($\beta=0$).
The distribution for $N=4$ independent eigenvalues are given by
\begin{align}
P_{0}^{4}(r) & =\frac{3}{2\pi\gamma}\left(2\sqrt{3}-\frac{2+r}{\sqrt{d_{344}}}-\frac{1+2r}{\sqrt{d_{443}}}\right).\label{eq_P^4_0}
\end{align}

Mixtures where one of the component only has one eigenvalue gives identical results to setting the $\beta=0$ for that component, giving $P_{\beta,\alpha}^{2,1}=P_{\beta,0}^{2,1}$.
The surmises for not setup are thus given by
\begin{align}
  P_{1,0}^{2,1}(r) & =\frac{3}{4}\frac{\delta}{\gamma^{3/2}}\label{eq_P^(2,1)_(1,0)}\\
  P_{2,0}^{2,1}(r) & =\frac{3\sqrt{3}}{2\pi\gamma}\label{eq_P^(2,1)_(2,0)}\\
  P_{3,0}^{2,1}(r) & =\frac{27\left(2r^{3}+3r^{2}+3r+2\right)}{64\gamma^{5/2}}\label{eq_P^(2,1)_(3,0)}\\
  P_{4,0}^{2,1}(r) & =\frac{3\sqrt{3}}{2\pi\gamma},\label{eq_P^(2,1)_(4,0)}
\end{align}
where we note that $P_{2,0}^{2,1}=P_{4,0}^{2,1}(r)$.
For $N=3$ the results are
\begin{align}
  P_{1,0}^{3,1}(r)  =&\frac{\delta^{2}}{\gamma^{2}}\frac{E_{1,1}^{3,1}(r)}{\sqrt{2}\pi d_{443}^{2}d_{344}^{2}}\label{eq_P^(3,1)_(1,0)}\\
  &+\frac{\left(r^{2}+10r+1\right)\delta}{4\pi\gamma^{5/2}}\left[\pi-T(r)-T\left(\frac{1}{r}\right)\right]\nonumber\\
  P_{2,0}^{3,1}(r)=&\frac{\sqrt{3}E_{2,0}^{3,1}(r)}{2\pi\delta^{4}}
  -\frac{F_{2,0}^{3,1}(r)}{4\pi\delta^{4}d_{344}^{7/2}}
  -\frac{G_{2,0}^{3,1}(r)}{4\pi\delta^{4}d_{443}^{7/2}}\label{eq_P^(3,1)_(2,0)}\\
  P_{0,3}^{1,3}\left(r\right)= & \frac{3E_{0,3}^{1,3}\left(r\right)\delta}{224\pi\gamma^{11/2}}\left[\pi-T\left(r\right)-T\left(\frac{1}{r}\right)\right]\nonumber\\
 & +\frac{d^{2}\tilde{E}_{0,3}^{1,3}\left(r\right)}{280\sqrt{2}\pi\gamma^{5}d_{344}^{5}d_{443}^{5}}\label{eq_P^(3,1)_(3,0)}
\end{align}
with
\begin{align*}
E_{1,0}^{3,1}(r) & =312r^{8}+708r^{7}+1142r^{6}+969r^{5}\\
 & +998r^{4}+969r^{3}+1142r^{2}+708r+312.
\end{align*}
and 
\begin{align*}
  E_{2,0}^{3,1}(r)=&4r^{6}+12r^{5}+51r^{4}+82r^{3}+51r^{2}+12r+4\\
  F_{2,0}^{3,1}(r)
   =&-36r^{13}-276r^{12}+677r^{11}+10816r^{10}\\
  & +45017r^{9}+110550r^{8}+185820r^{7}\\
  & +225768r^{6}+201768r^{5}+131664r^{4}\\
  &+61248r^{3}+19456r^{2}+4160r+640\\
  G_{2,0}^{3,1}(r)
   =&640r^{13}+4160r^{12}+19456r^{11}+61248r^{10}\\
  & +131664r^{9}+201768r^{8}+225768r^{7}\\
  & +185820r^{6}+110550r^{5}+45017r^{4}\\
  &+10816r^{3}+677r^{2}-276r-36.
\end{align*}

\begin{align*}
E_{0,3}^{1,3}\left(r\right)= & 44r^{8}+154r^{7}+175r^{6}+1476r^{5}\\
 & +2710r^{4}+1476r^{3}+175r^{2}+154r+44
\end{align*}

\begin{widetext}
\begin{align*}
  \tilde{E}_{0,3}^{1,3}\left(r\right)  =
  & 133318656r^{26}+1849393152r^{25}+14056292352r^{24}+71482931456r^{23}+264844710656r^{22}\\
  & +748372542976r^{21} + 1652680538624r^{20} + 2871205333152r^{19} + 3859106493224r^{18} +3743668969440r^{17} \\
  & +1905039479728r^{16}-1307897322555r^{15} -4438665127726r^{14}-5742939216774r^{13}-4438665127726r^{12}\\
  & -1307897322555r^{11} + 1905039479728r^{10} + 3743668969440r^{9} + 3859106493224r^{8}+2871205333152r^{7} \\
  & +1652680538624r^{6} +748372542976r^{5}+264844710656r^{4}+71482931456r^{3} \\
  & +14056292352r^{2}+1849393152r+133318656
\end{align*}
\end{widetext}

The integrals needed to reach higher $\beta$ include increasing amount of terms and are left for the future.

There are no surmises of the form $P_{\beta,0,0}^{N,1,1}$ since if several components have only one eigenvalue, they can be modeled as one single component with $\beta=0$, giving  $P_{\beta,0,0}^{N,1,1}=P_{\beta,0}^{N,2}$
Thus, the next set of surmises is $P_{\beta,0}^{2,2}$ given by
\begin{align}
P_{0,1}^{2,2}(r)= & \frac{-3\delta}{4\pi}\frac{T(r)+T\left(\frac{1}{r}\right)+\pi}{\gamma^{3/2}}\nonumber \\
 & +\frac{3}{\sqrt{2}\pi}\frac{6r^{4}+9r^{3}+14r^{2}+9r+6}{\gamma d_{344}d_{443}}\label{eq_P^(2,2)_(0,1)}\\
P_{0,2}^{2,2}(r)  =&\frac{3\sqrt{3}}{\pi\gamma}-\frac{3}{2\pi\gamma}\left(\frac{r+2}{d_{344}^{\frac{1}{2}}}+\frac{2r+1}{d_{443}^{\frac{1}{2}}}\right)\label{eq_P^(2,2)_(0,2)}\\
P_{0,3}^{2,2}(r)= & \frac{27}{64}\frac{2r^{3}+3r^{2}+3r+2}{\gamma^{5/2}}\nonumber \\
& -\frac{3}{32\pi}\frac{F_{0,3}^{2,2}(r)T(r)+G_{0,3}^{2,2}(r)T\left(\frac{1}{r}\right)}{\gamma^{5/2}}\nonumber \\
 & +\frac{E_{0,3}^{2,2}}{4\sqrt{2}\pi\gamma^{2}d_{344}^{2}d_{443}^{2}}\label{eq_P^(2,2)_(0,3)}\\
P_{0,4}^{2,2}(r)  =&  \frac{3 \sqrt{3}}{\pi\gamma}-\frac{F_{0,4}^{2,2}(r)}{6 \pi  \gamma d_{344}^{5/2}}-\frac{G_{0,4}^{2,2}(r)}{6 \pi  \gamma d_{443}^{5/2}}\label{eq_P^(2,2)_(0,4)}
\end{align}

The auxiliary functions for $\beta=3$ and $\beta=4$ are given by 
\begin{align*}
F_{0,3}^{2,2}(r)= & 10r^{3}+15r^{2}+12r+8\\
G_{0,3}^{2,2}(r)= & 8r^{3}+12r^{2}+15r+10\\
E_{0,3}^{2,2}(r)= & 816r^{10}+3960r^{9}+12012r^{8}+24290r^{7}\\
 & +36876r^{6}+41927r^{5}+36876r^{4}+24290r^{3}\\
 & +12012r^{2}+3960r+816
\end{align*}
and
\begin{align*}
F_{0,4}^{2,2}(r) & =69r^{5}+350r^{4}+788r^{3}+1032r^{2}+760r+304\\
G_{0,4}^{2,2}(r) & =304r^{5}+760r^{4}+1032r^{3}+788r^{2}+350r+69.
\end{align*}

\subsection{Equal $\beta$ mixtures $P_{\beta,\beta}^{2,2}$}
In this subsection we list the surmises that figure in the main text, and are on the form $P_{\beta,\beta}^{2,2}$.
These are

\begin{align}
P_{1,1}^{2,2}(r)  =&\frac{\delta}{4\gamma^{2}}\left(\frac{F_{1,1}^{2,2}(r)}{d_{443}^{3/2}}+r\frac{G_{1,1}^{2,2}(r)}{d_{344}^{3/2}}\right)\label{eq_P^(2,2)_(1,1)}\\
P_{2,2}^{2,2}(r)= & \frac{3}{\pi\gamma}\left(\sqrt{3}-\frac{F_{2,2}^{2,2}(r)}{4d_{443}^{5/2}}-\frac{G_{2,2}^{2,2}(r)}{4d_{344}^{5/2}}\right).\label{eq_P^(2,2)_(2,2)}\\
P_{3,3}^{2,2}(r)= & \frac{3\delta}{64\gamma^{4}}\left(\frac{F_{3,3}^{2,2}(r)}{d_{344}^{7/2}}+\frac{G_{3,3}^{2,2}(r)}{d_{443}^{7/2}}\right).\label{eq_P^(2,2)_(3,3)}\\
  P_{4,4}^{2,2}(r)= & \frac{1}{\gamma^{4}}\frac{3}{2\pi}\left(\frac{2E_{4,4}^{2,2}(r)}{3\sqrt{3}}-\frac{\left(2r+1\right)F_{4,4}^{2,2}(r)}{d_{443}^{9/2}}\right.\label{eq_P^(2,2)_(4,4)}\\
  & \phantom{aaaaa}\left.-\frac{\left(2+r\right)G_{4,4}^{2,2}(r)}{d_{344}^{9/2}}\right).\nonumber 
\end{align}

The auxiliary functions are given by

\begin{align*}
F_{1,1}^{2,2}(r) & =16r^{3}+34r^{2}+31r+15\\
G_{1,1}^{2,2}(r) & =15r^{3}+31r^{2}+34r+16,
\end{align*}
\begin{align*}
F_{2,2}^{2,2}(r) & =\left(2r+1\right)\left(\left(4r^{2}+4r+7\right)^{2}-7\right)\\
G_{2,2}^{2,2}(r) & =\left(r+2\right)\left(\left(7r^{2}+4r+4\right)^{2}-7r^{4}\right)
\end{align*}
\begin{align*}
F_{3,3}^{2,2}(r) & =918r^{12}+6120r^{11}+22107r^{10}+53227r^{9}\\
 & +92358r^{8}+119122r^{7}+114220r^{6}+79416r^{5}\\
 & +36816r^{4}+8528r^{3}-1248r^{2}-1408r-256,\\
G_{3,3}^{2,2}(r) & =-256r^{12}-1408r^{11}-1248r^{10}+8528r^{9}\\
 & +36816r^{8}+79416r^{7}+114220r^{6}+119122r^{5}\\
 & +92358r^{4}+53227r^{3}+22107r^{2}+6120r+918.
\end{align*}
\begin{align*}
E_{4,4}^{2,2}(r) & =11r^{6}+33r^{5}+39r^{4}+23r^{3}+39r^{2}+33r+11\\
F_{4,4}^{2,2}(r) & =96r^{14}+672r^{13}+2144r^{12}+4128r^{11}\\
 & +6712r^{10}+11736r^{9}+22172r^{8}+36224r^{7}\\
 & +46551r^{6}+45581r^{5}+33883r^{4}+18731r^{3}\\
 & +7467r^{2}+1953r+279\\
G_{4,4}^{2,2}(r) & =279r^{14}+1953r^{13}+7467r^{12}+18731r^{11}\\
 & +33883r^{10}+45581r^{9}+46551r^{8}+36224r^{7}\\
 & +22172r^{6}+11736r^{5}+6712r^{4}+4128r^{3}\\
 & +2144r^{2}+672r+96.
\end{align*}

\subsection{Distributions with mixed $\beta$}
Mixtures with 2+2 eigenvalues also admit a series of surmises with mixed values of beta.
Below we list these 6 combinations for $1\leq\beta\leq4$:

\paragraph{Surmise $P_{1,2}^{2,2}(r)$:}
\begin{align}
P_{1,2}^{2,2}(r) & =\frac{22r^{3}+53r^{2}+53r+22}{32\gamma^{5/2}}\label{eq_P^(2,2)_(1,2)}\\
 & -\frac{F_{1,2}^{2,2}(r)T(r)+G_{1,2}^{2,2}(r)T\left(\frac{1}{r}\right)}{16\pi\gamma^{5/2}}\nonumber \\
 & +\frac{E_{1,2}^{2,2}(r)}{2\sqrt{2}\pi\gamma^{2}d_{344}^{2}d_{443}^{2}}\nonumber 
\end{align}
\begin{align*}
F_{1,2}^{2,2}(r) & =2r^{3}+13r^{2}+40r+20\\
G_{1,2}^{2,2}(r) & =20r^{3}+40r^{2}+13r+2\\
E_{1,2}^{2,2}(r) & =384r^{10}+2160r^{9}+6896r^{8}+14200r^{7}\\
 & +21616r^{6}+24559r^{5}+21616r^{4}\\
 & +14200r^{3}+6896r^{2}+2160r+384.
\end{align*}

\paragraph{Surmise $P_{1,3}^{2,2}(r)$:}
\begin{equation}
P_{1,3}^{2,2}(r)=\frac{1}{8\gamma^{3}}\left(\frac{F_{1,3}^{2,2}(r)}{d_{344}^{5/2}}+\frac{G_{1,3}^{2,2}(r)}{d_{443}^{5/2}}\right)\label{eq_P^(2,2)_(1,3)}
\end{equation}
\begin{align*}
F_{1,3}^{2,2}(r) & =96r^{9}+533r^{8}+1518r^{7}+2719r^{6}\\
 & +3306r^{5}+2748r^{4}+1488r^{3}+448r^{2}+24r-16\\
G_{1,3}^{2,2}(r) & =-16r^{9}+24r^{8}+448r^{7}+1488r^{6}\\
 & +2748r^{5}+3306r^{4}+2719r^{3}+1518r^{2}+533r+96
\end{align*}

\paragraph{Surmise $P_{1,4}^{2,2}(r)$:}
\begin{align}
P_{1,4}^{2,2}(r)= & \frac{E_{1,4}^{2,2}(r)}{128\gamma^{7/2}}+\frac{F_{1,4}^{2,2}T(r)+G_{1,4}^{2,2}T\left(\frac{1}{r}\right)}{64\pi\gamma^{7/2}}\nonumber \\
 & +\frac{\tilde{E}_{1,4}^{2,2}(r)}{48\sqrt{2}\pi\gamma^{3}d_{344}^{3}d_{443}^{3}}\label{eq_P^(2,2)_(1,4)}
\end{align}
\begin{align*}
E_{1,4}^{2,2}(r) & =78r^{5}+307r^{4}+566r^{3}+566r^{2}+307r+78\\
F_{1,4}^{2,2}(r) & =6r^{5}-41r^{4}-238r^{3}-328r^{2}-266r-84\\
G_{1,4}^{2,2}(r) & =-84r^{5}-266r^{4}-328r^{3}-238r^{2}-41r+6
\end{align*}
\begin{align*}
\tilde{E}_{1,4}^{2,2}(r) & =113664r^{16}+1046720r^{15}+5063040r^{14}\\
 & +16406704r^{13}+39749568r^{12}+75927300r^{11}\\
 & +118307400r^{10}+153231453r^{9}+166874784r^{8}\\
 & +153231453r^{7}+118307400r^{6}+75927300r^{5}\\
 & +39749568r^{4}+16406704r^{3}+5063040r^{2}\\
 & +1046720r+113664
\end{align*}

\paragraph{Surmise $P_{2,3}^{2,2}(r)$:}
\begin{align}
P_{2,3}^{2,2}(r) & =\frac{9E_{2,3}^{2,2}(r)}{256\gamma^{7/2}}-\frac{3\left(F_{2,3}^{2,2}(r)T(r)+G_{2,3}^{2,2}(r)T\left(\frac{1}{r}\right)\right)}{128\pi\gamma^{7/2}}\nonumber \\
 & +\frac{\tilde{E}_{2,3}^{2,2}(r)}{32\sqrt{2}\pi\gamma^{3}d_{344}^{3}d_{443}^{3}}\label{eq_P^(2,2)_(2,3)}
\end{align}
\begin{align*}
E_{2,3}^{2,2}(r) & =26r^{5}+65r^{4}+74r^{3}+74r^{2}+65r+26\\
F_{2,3}^{2,2}(r) & =10r^{5}+25r^{4}+82r^{3}+140r^{2}+170r+68\\
G_{2,3}^{2,2}(r) & =68r^{5}+170r^{4}+140r^{3}+82r^{2}+25r+10\\
\tilde{E}_{2,3}^{2,2}(r) & =50688r^{16}+464832r^{15}+2619648r^{14}\\
 & +10088176r^{13}+28593440r^{12}+62047380r^{11}\\
 & +106098928r^{10}+145418249r^{9}+161380392r^{8}\\
 & +145418249r^{7}+106098928r^{6}+62047380r^{5}\\
 & +28593440r^{4}+10088176r^{3}+2619648r^{2}\\
 & +464832r+50688
\end{align*}

\paragraph{Surmise $P_{2,4}^{2,2}(r)$:}
\begin{align}
P_{2,4}^{2,2}(r) & =\frac{E_{2,4}^{2,2}(r)}{4\sqrt{3}\pi\gamma^{4}}-\frac{F_{2,4}^{2,2}(r)}{24\pi\gamma^{4}d_{344}^{7/2}}-\frac{G_{2,4}^{2,2}(r)}{24\pi\gamma^{4}d_{443}^{7/2}}\label{eq_P^(2,2)_(2,4)}
\end{align}
\begin{align*}
E_{2,4}^{2,2}(r) & =38r^{6}+114r^{5}+201r^{4}+212r^{3}\\
 & +201r^{2}+114r+38\\
F_{2,4}^{2,2}(r) & =2682r^{13}+20562r^{12}+79123r^{11}+200738r^{10}\\
 & +368767r^{9}+518820r^{8}+574926r^{7}+516036r^{6}\\
 & +380856r^{5}+235584r^{4}+119856r^{3}+48416r^{2}\\
 & +13312r+2048\\
G_{2,4}^{2,2}(r) & =2048r^{13}+13312r^{12}+48416r^{11}+119856r^{10}\\
 & +235584r^{9}+380856r^{8}+516036r^{7}+574926r^{6}\\
 & +518820r^{5}+368767r^{4}+200738r^{3}+79123r^{2}\\
 & +20562r+2682
\end{align*}

\paragraph{Surmise $P_{3,4}^{2,2}(r)$:}
\begin{align}
P_{3,4}^{2,2}(r) & =\frac{E_{3,4}^{2,2}(r)}{1024\gamma^{9/2}}-\frac{F_{3,4}^{2,2}(r)T(r)+G_{3,4}^{2,2}(r)T\left(\frac{1}{r}\right)}{512\pi\gamma^{9/2}}\nonumber \\
 & +\frac{\tilde{E}_{3,4}^{2,2}(r)}{384\sqrt{2}\pi\gamma^{4}d_{443}^{3}d_{443}^{4}}\label{eq_P^(2,2)_(3,4)}
\end{align}
\begin{align*}
E_{3,4}^{2,2}(r) & =934r^{7}+3269r^{6}+5775r^{5}+7977r^{4}\\
 & +7977r^{3}+5775r^{2}+3269r+934\\
F_{3,4}^{2,2}(r) & =2r^{7}+7r^{6}+795r^{5}+2826r^{4}+5151r^{3}\\
 & +4980r^{2}+3262r+932\\
G_{3,4}^{2,2}(r) & =932r^{7}+3262r^{6}+4980r^{5}+5151r^{4}\\
 & +2826r^{3}+795r^{2}+7r+2
\end{align*}
\begin{align*}
 & \tilde{E}_{3,4}^{2,2}(r)=\\
 & =10285056r^{22}+126869760r^{21}+832050432r^{20}\\
 & +3751871232r^{19}+12895150912r^{18}+35584672928r^{17}\\
 & +81305771632r^{16}+156811861200r^{15}+258577524828r^{14}\\
 & +367643491941r^{13}+453161735140r^{12}+485718539552r^{11}\\
 & +453161735140r^{10}+367643491941r^{9}+258577524828r^{8}\\
 & +156811861200r^{7}+81305771632r^{6}+35584672928r^{5}\\
 & +12895150912r^{4}+3751871232r^{3}+832050432r^{2}\\
 & +126869760r+10285056
\end{align*}

\section{Summary and Discussion}

This work used computer-generated algebra to compute gap ratios for all mixed Wigner surmises with $N=3$ and $N=4$ and $0\geq\beta\geq 4$.
Due to the memory constraints and evaluation times, we did not push the calculations to larger mixed ensembles. 
There is, however, no principal problem with looking at a larger number of eigenvalues.
With improved algorithms, undoubtedly, larger ensembles of eigenvalue can be targeted.

We wish to point out that the numerical approach combined Julia and Mathematica code.
The Julia controlled the flow of logic and algorithms,
whereas Mathematica evaluated the algebraic expressions and performed certain integrals.

As for the analytical results, we hope that these will prove helpful as a simple diagnostic for mixed distributions.
More elaborate diagnostics are also possible by considering higher-order statistics \cite{Tekur2018}, and this work adds to that toolbox. 

We find it a lucky coincidence that the $P_{3,3}^{2,2}$ surmise gives such a good approximation for the \TwoGSE~statistic,
and it seems $P_{3,3}^{3,3}$ would be an almost perfect match.
We leave it as an open question if the approximation is also suitable for higher-order correlations.
While writing this manuscript, we became aware of the work in Ref.~\onlinecite{Giraud2020},
where also the gap-ratio statistics of independent blocks were considered. However,
that work had a slightly different focus, and thus the analytic results are somewhat different.

The reader will note that this work is only concerned with $r$-statistics and that no results for $s$-statistics are listed.
The main reason for this is the problem that $s$-statistics has with unfolding, but there is also a technical aspect worth pointing out.
For many of the surmises investigated here, the techniques to compute the $r$-statistic do not work to calculate the $s$-statistic.
We thus leave these for future investigation.

\section*{Acknowledgments}

The author thanks Masud Haque and Lars Fritz for useful discussions.
This work made use of Mathematica and the MathLink and MathLinkExtras packages for Julia.

This work is part of the D-ITP consortium, a program of the Netherlands Organisation for Scientific Research (NWO) that is funded by the Dutch Ministry of Education, Culture and Science (OCW).

\begin{figure*}
\begin{centering}
\includegraphics[width=1\linewidth]{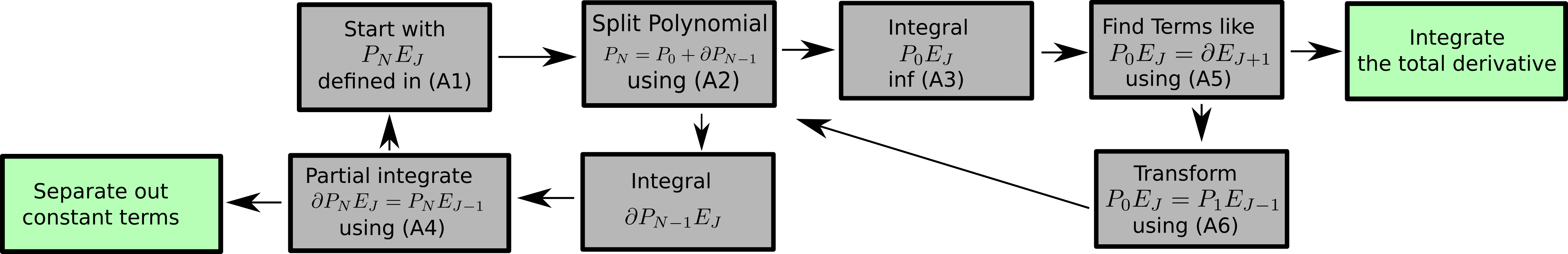}
\par\end{centering}
\caption{Flowchart of calculation\label{fig_Flowchart-of-calculation} described
in Appendix \ref{app_Recursive-integration}.}
\end{figure*}

\bibliography{Bibfile}

\appendix

\section{A recursive integration scheme\label{app_Recursive-integration}}

In this appendix, we describe schematically the steps needed to compute the distributions listed in section \ref{sec:Analytical-Results}.
In the course of computing the nested integral \eqref{eq_integral_term}, we will encounter a few different types of specific integrals.
The calculation relies on recursion and the flow is schematically illustrated in Figure \ref{fig_Flowchart-of-calculation}.
The first is the integral over the $\delta$-function, which is trivial, and we will not further mention it.

The second typical integral is of the form 
\begin{equation}
\int P_NE_J=\int_{0}^{\infty}dx\,F_{N}(x)e^{-f(x)}\prod_{j=1}^{J}\text{erf}\left(p_{j}(x)\right)\label{eq_TempInt_N,J}
\end{equation}
where $F_{N}(x)$ is a polynomial in $x$ of degree $N$,
$f(x)$ is a second order polynomial in $x$ and $p_{j}(x)$ are a set of $J$ first order polynomials in $x$.

The first step to integrate \eqref{eq_TempInt_N,J} is to rewrite the first two factors of the integrand as
\begin{equation}
F_N(x)e^{-f(x)}=F_0e^{-f(x)}+\frac{\partial}{\partial x}\left[F_{N-1}(x)e^{-f(x)}\right].\label{eq_Poly_split}
\end{equation}
Here $F_0$ is a constant and $F_{N-1}(x)$ is a polynomial one degree lower than $F_N(x)$.
Schematically we have $P_N=P_0 + \partial P_{N-1}$.
Inserting \eqref{eq_Poly_split} into \eqref{eq_TempInt_N,J}, one obtains two terms as
\[ \int P_NE_J = \int P_0E_J +\int (\partial P_{N-1})E_J.\]
The first, $\int P_0E_J$, has no polynomial part and is given by 
\begin{equation}
\int P_0E_J=F_0\int_0^{\infty}dx\,e^{-f(x)}\prod_{j=1}^{J}\text{erf}\left(p_{j}(x)\right)\label{eq_TempInt_0,J}.
\end{equation}
We leave this term as it is for the time being.

The second piece $\int (\partial P_{N-1})E_J$, contains an explicit derivative and can be approached by partial integration as
\begin{widetext}
\begin{align}
\int (\partial P_{N-1})E_J & =\int_{0}^{\infty}dx\,\frac{\partial}{\partial x}\left[F_{N-1}(x)e^{-f(x)}\right]\prod_{j=1}^{J}\text{erf}\left(p_{j}(x)\right)\label{eq_Spil_Part_Int}\\
 & =\left[F_{N-1}(x)e^{-f(x)}\prod_{j=1}^{J}\text{erf}\left(p_{j}(x)\right)\right]_{0}^{\infty}-\int_{0}^{\infty}dx\,\left[F_{N-1}(x)e^{-f(x)}\right]\frac{\partial}{\partial x}\left(\prod_{j=1}^{J}\text{erf}\left(p_{j}(x)\right)\right)\nonumber.
\end{align}
\end{widetext}
The first term can be evaluated directly, and the second term (after acting with the derivatives) is again of the form \eqref{eq_TempInt_N,J}, but now with $N\to N-1$ and $J\to J-1$.
By recursion, which stops when $N=0$, the only unintegrated terms left will be of the form $\int P_0E_J$ given Eqn.~\eqref{eq_TempInt_0,J}. 

The integral $\int P_0E_J$ is integrated in two steps.
First we search for terms that can be combined to form products of $\text{erf}$-functions $\prod_{j=1}^{J+1}\text{erf}\left(p_{j}(x)\right)$.
For this we use that 
\begin{equation}
  \frac{\partial}{\partial x}\prod_{j=1}^{J+1}\text{erf}\left(p_j(x)\right)=
  \sum_{k=1}^{J+1}\left(\frac{\partial p_j}{\partial x}\right)e^{-p_j^2(x)}\prod_{k\neq j=1}^{J+1}\text{erf}\left(p_j(x)\right)\label{eq_Integrate_PrductErfs}
\end{equation}
is a sum of terms on the form of $P_0E_J$.
If the total derivative $\partial(E_{J+1})=P_0E_J$ can be identified, the integral is trivially evaluated and put aside.

There may still be terms of the form $\int P_0E_J$ that cannot be integrated by making use of \eqref{eq_Integrate_PrductErfs}.
To make progress, we represent the error function of $x$ as an integral over an auxiliary variable $t$:
\begin{equation}
\text{erf}(x)=x\sqrt{\frac{2}{\pi}}\int_{0}^{1}dt\,e^{-x^{2}t^{2}}.\label{eq_Erf_rep}
\end{equation}
With this transformation we can represent $\int P_0E_J$ as an integral on the form $\int P_1E_{J-1}$ and repeat the steps following equation \eqref{eq_Poly_split},
until all error functions have been integrated (or transformed away).

Using the auxiliary variables $t$, we can perform all the integrals,
with the price that we now must integrate over the additional variables $t_i$ at the end.
Fortunately, the integrals that we consider in this work can always be performed with the help of Euler substitution.

Once all the computations have been performed, the total probability distribution is given by the sum of all the parts
\begin{align*}
P(r) & =\frac{\sum_{k,\sigma}P_{k,\sigma}(r)}{\sum_{k,\sigma}Z_{\sigma}},
\end{align*}
where $Z_{\sigma}$ are the normalization constants of $P_{\sigma}\left(g\right)$.

\end{document}